\documentclass[aps,pra,twocolumn,showpacs,superscriptaddress]{revtex4-1}

\usepackage{dcolumn}   
\usepackage{amssymb,amsthm,amsmath,amsfonts}
\usepackage{color, hyperref,graphicx}
\usepackage{ulem}
\usepackage{rotating}

\begin{document}
 
\title{Full Multipartite Entanglement of Frequency-Comb Gaussian States}

\author{S. Gerke}\email{stefan.gerke@uni-rostock.de}
\affiliation{Arbeitsgruppe Theoretische Quantenoptik, Institut f\"ur Physik, Universit\"at Rostock, D-18051 Rostock, Germany} 
\author{J. Sperling}
\affiliation{Arbeitsgruppe Theoretische Quantenoptik, Institut f\"ur Physik, Universit\"at Rostock, D-18051 Rostock, Germany} 
\author{W. Vogel}
\affiliation{Arbeitsgruppe Theoretische Quantenoptik, Institut f\"ur Physik, Universit\"at Rostock, D-18051 Rostock, Germany} 
\author{Y. Cai}
\affiliation{Laboratoire Kastler Brossel, Sorbonne Universit\'es - UPMC, \'Ecole Normale Sup\'erieure, Coll\`ege de France, CNRS; 4 place Jussieu, 75252 Paris, France}
\author{J. Roslund}
\affiliation{Laboratoire Kastler Brossel, Sorbonne Universit\'es - UPMC, \'Ecole Normale Sup\'erieure, Coll\`ege de France, CNRS; 4 place Jussieu, 75252 Paris, France}
\author{N. Treps}
\affiliation{Laboratoire Kastler Brossel, Sorbonne Universit\'es - UPMC, \'Ecole Normale Sup\'erieure, Coll\`ege de France, CNRS; 4 place Jussieu, 75252 Paris, France}
\author{C. Fabre}
\affiliation{Laboratoire Kastler Brossel, Sorbonne Universit\'es - UPMC, \'Ecole Normale Sup\'erieure, Coll\`ege de France, CNRS; 4 place Jussieu, 75252 Paris, France}
\date{\today}

\begin{abstract}
	An analysis is conducted of the multipartite entanglement for Gaussian states generated by the parametric down-conversion of a femtosecond frequency comb.
	Using a recently introduced method for constructing optimal entanglement criteria, a family of tests is formulated for mode decompositions that extends beyond the traditional bipartition analyses.
	A numerical optimization over this family is performed to achieve maximal significance of entanglement verification.
	For experimentally prepared 4-, 6-, and 10-mode states, full entanglement is certified for all of the 14, 202, and 115\,974 possible nontrivial partitions, respectively.
\end{abstract}
\maketitle

\section{Introduction}
One of the most fundamental concepts in quantum physics is entanglement~\cite{EPR35,SB35,SD36}.
This property plays a central role in a host of quantum technologies, including metrology, imaging, communication, and quantum information processing~\cite{NielsenChuang,HHHH09,GT09}.
Protocols in each of these domains rely upon the existence of nonclassical correlations among a multitude of subsystems within a multimode state~\cite{HV13,LM13,GK14}.
As such, reliable, readily implementable, and versatile means of characterizing entanglement are essential for assessing the utility of certain states as well as understanding the fundamental physics underlying quantum interactions.

One method for identifying entanglement is formulated in terms of positive -- but not completely positive -- maps.
The most prominent example of such a map is the partial transposition (PT)~\cite{Peres96}.
For bipartite Gaussian states, which are completely characterized by the covariance matrix, it has been shown that the PT criterion is necessary and sufficient to identify entanglement~\cite{Simon00,DGCZ00}.
In the multipartite case, however, the PT criteria can only diagnose entanglement among bipartitions.
Moreover, bound entangled Gaussian states are known to exist whose entanglement cannot be detected with the PT criterion.
Such states have been formulated in theory and also realized in experiments~\cite{WW01,DSHPES11}.
Additionally, a number of moment-based entanglement probes have been successfully deployed to characterize entanglement, e.g., Refs.~\cite{SV05,SV06,MPHH09,GHGE07,AB05,LF03,SSV13,HE06,Serafini06,HZ06}.

These criteria have been enormously successful at experimentally diagnosing entanglement among various beams~\cite{STJZXP07,YULF08} or among different parties of a multimode beam~\cite{PMSBF11,AMJHTLB12}.
Alternatively, several studies have acquired the covariance matrix for a multidimensional state, which enables implementation of the PT criterion as a means for examining the nonclassical correlations among multiple beams~\cite{CBCVMN09,VSEHFS13}.
In each of these situations, however, the employed methods restrict multipartite dynamics to the set of all possible bipartite state divisions. 

Another well-established method for identifying entanglement is formulated in terms of entanglement witnesses~\cite{HHH96,HHH01}.
In particular, the {\sl separability eigenvalue equations} have recently been introduced as a method for constructing optimal witnesses~\cite{SV09,SV13}.
The solutions of these coupled equations yield powerful entanglement assessments not only for bipartite divisions but also for high-order multipartite divisions of discrete and continuous variable quantum systems.

This Letter formulates entanglement conditions for multimode Gaussian states and subsequently demonstrates their application on an experimentally realized quantum ultrafast frequency comb.
This quantum state, which is generated by the parametric down-conversion of a classical frequency comb, was recently shown to exhibit bipartite entanglement among its underlying frequency bands~\cite{RMJFT14}. 
The covariance matrix for this high-dimensional quantum object has been measured, which renders it a unique test bed for exploring novel multipartite entanglement metrics.
Importantly, we will show in this Letter that the criteria developed from the separability eigenvalue equations are able to examine nonclassical aspects of the frequency comb not feasible with strictly bipartite methods. 
Within this class of criteria, the significance of the verified entanglement is optimized with a genetic algorithm, which allows us to fully verify the entanglement present in highly complex multiparty quantum systems.
For the 10-mode system considered here, entanglement is certified for each of the 115\,974 possible nontrivial mode partitions.

\section{Gaussian states and mode decompositions}
Gaussian states are described by a Gaussian characteristic function on a multimode phase space (for an introduction, see, e.g., Ref.~\cite{AI07}).
The amplitude and phase quadratures of individual modes are denoted by $\hat{x}_k$ and $\hat{p}_k$, respectively, and a vector of quadratures is defined as
\begin{equation}\label{eq:quadratures}
	\hat{\xi}=(\hat{x}_1,\ldots,\hat{x}_N,\hat{p}_1,\ldots,\hat{p}_N)^{\rm{T}}.
\end{equation}
The  covariance matrix $C$ is then specified by its entries
\begin{equation}
	C^{ij}=\frac{1}{2}\langle\hat{\xi}_i\hat{\xi}_j+\hat{\xi}_j\hat{\xi}_i\rangle-\langle\hat{\xi}_i\rangle\langle\hat{\xi}_j\rangle.
\end{equation}
First-order moments are irrelevant for entanglement since local unitary displacement operations may be applied to yield $\langle \hat\xi\rangle=0$.
Thus, without loss of generality, we can assume that all of the information for a Gaussian state is contained in its second-order moments. 

The initial set of $N$ orthonormal modes, on which the multimode quantum state is defined, can be decomposed into many different partitions, each one distributing the $N$ modes in $K$ different and complementary subsystems $\mathcal{I}_1{:}\cdots{:}\mathcal{I}_K$, with $K$ being any integer between $1$ and $N$.
A quantum state is considered entangled with respect to a given mode partitioning if one is not able to write it as a statistical mixture of product density matrices $\hat{\varrho}_1\otimes\cdots\otimes\hat{\varrho}_K$, where $\hat{\varrho}_j$ describes a quantum state in subsystem $\mathcal{I}_j$ for $j=1,\dots,K$.
The case $K=2$ consists of $2^{N-1}-1$ mode bipartitions, which are the only ones addressed by the PT criterion.
However, even if entanglement does not exist among certain bipartitions, it may be present in higher-order partitions, i.e., $K>2$.
Considering that the total number of state partitions is given by the Bell number and increases rapidly as a function of $N$~\cite{BT10}, the PT criterion addresses only a very small subset of the rich variety of possible partitionings. 

\section{Optimal entanglement tests}
The multipartite entanglement of a quantum state $\hat \rho$ may be probed with the use of a general Hermitian operator $\hat{L}$~\cite{SV13}.
In particular, the state under question is entangled with respect to a given $K$ partition if and only if it may be shown that 
\begin{equation}\label{eq:entanglement}
	{\rm tr}(\hat{L}\hat{\rho})<g^{\min}_{\mathcal{I}_1{:}\cdots{:}\mathcal{I}_K},
\end{equation}
where $g^{\min}_{\mathcal{I}_1{:}\cdots{:}\mathcal{I}_K}$ is the minimum expectation value of $\hat{L}$  among all separable states of the $K$ partition.
It was established in Ref.~\cite{SV13} that this minimization problem can be solved with a set of coupled eigenvalue equations, denoted as separability eigenvalue equations.
The resulting minimal separability eigenvalue is identical to $g^{\min}_{\mathcal{I}_1{:}\cdots{:}\mathcal{I}_K}$.

The most general form of the operator $\hat{L}$ for continuous variable Gaussian states is given as $\hat{L} = \sum_{i,j} \left( M_{xx}^{ji} \hat{x}_{i} \hat{x}_{j} + M_{px}^{ji} \hat{p}_{i} \hat{x}_{j} + M_{xp}^{ji} \hat{x}_{i} \hat{p}_{j} + M_{pp}^{ji} \hat{p}_{i} \hat{p}_{j} \right)$ in which the coefficients of $M$ are freely adjustable.
Accordingly, attention may be restricted to the state's covariance matrix.
Correlations between the amplitude and phase quadratures are negligible for the presently studied states, which allows the test operator $\hat{L}$ to be cast as
\begin{equation}\label{Eq:4}
	\hat L={\rm Tr}(M\hat\xi\hat\xi^{\rm T}), \text{ with }
	M=\begin{pmatrix}
	M_{xx}	&	0	\\	0	&	M_{pp}
	\end{pmatrix}=M^{\rm T}>0,
\end{equation}
where $M_{xx}$ and $M_{pp}$ are coefficient matrices of the same dimensionality as the corresponding state covariance matrix, and the indices $xx$ and $pp$ refer to amplitude-amplitude and phase-phase correlations, respectively.
The expectation value of this test operator readily follows and is written as~\cite{CommentTrace}
\begin{equation}
	\langle\hat{L}\rangle={\rm tr}(\hat L\hat \rho)={\rm Tr}(MC).
\end{equation}
Likewise, the minimal separability eigenvalue $g_{\mathcal{I}_1{:}\cdots{:}\mathcal{I}_K}^{\min}$ for operators of this form has been derived in Ref.~\cite{SV13} and reads as
\begin{equation}
	g_{\mathcal{I}_1{:}\cdots{:}\mathcal{I}_K}^{\min}=\sum_{k=1}^K {\rm Tr}_{\mathcal I_k}\left[M_{pp,\mathcal I_k}^{1/2}M_{xx,\mathcal I_k}M_{pp,\mathcal I_k}^{1/2}\right]^{1/2},
\end{equation}
where $M_{\mathcal I_k}$ are the submatrices of $M$ that contain only the rows and columns of the modes within $\mathcal I_k$.
The solution for general Gaussian test operators is given in Ref.~\cite{supplement}.

A partition's entanglement is characterized in terms of its statistical significance $\Sigma$, which compares the difference between the expectation value $\langle\hat{L}\rangle$ and its separable bound $g_{\mathcal{I}_1{:}\cdots{:}\mathcal{I}_K}^{\min}$ to the experimental standard deviation $\sigma(L)$:
\begin{equation}\label{eq:Sigma}
	\Sigma=\frac{\langle \hat L\rangle -g_{\mathcal{I}_1{:}\cdots{:}\mathcal{I}_K}^{\min}}{\sigma(L)},
\end{equation}
which is the considered entanglement metric.
The experimental error $\sigma(L)$ is determined through error propagation of $\langle \hat{L} \rangle$ and yields 
\begin{equation}
	\sigma(L)=\sqrt{\sum_{i,j=1}^N([M_{xx}^{ij}]^2[\sigma(C_{xx}^{ji})]^2+[M_{pp}^{ij}]^2[\sigma(C_{pp}^{ji})]^2)},
\end{equation}
where $\sigma(C_{xx}^{ji})$ and $\sigma(C_{pp}^{ji})$ are the measured errors corresponding to the covariance elements $C_{xx}^{ji}$ and $C_{pp}^{ji}$, respectively.
A partition is considered to be entangled if $\Sigma<0$, and the statistical significance of its nonseparability is assessed with $|\Sigma|$.
The coefficient matrix $M$ may be freely tuned in order to maximize the significance of each partition, $\Sigma\to\Sigma_{\min}<0$.
This optimization is achieved with a \textsl{genetic algorithm} (see Refs.~\cite{GeneticRef,supplement} for details).

\section{Experimental realization}
Femtosecond frequency combs contain upwards of $\sim10^{5}$ individual frequency components, and the simultaneous down-conversion of all of these frequencies in a nonlinear crystal inserted in an optical cavity initiates a network of frequency correlations that extends across the width of the resultant comb~\cite{VPTF06}.
The laser source utilized to create the entangled comb is a titanium:sapphire mode-locked oscillator that delivers  $\sim 6\,\textrm{nm}$ FWHM pulses ($\sim 140\,\textrm{fs}$) centered at 795\,nm with a repetition rate of 76\,MHz.
This pulse train is frequency doubled, which serves to pump a below-threshold optical parametric oscillator (OPO) containing a 2\,mm BIBO crystal~\cite{ARCFFT14}.
The state exiting the OPO is analyzed with homodyne detection, in which the spectral composition of the local oscillator (LO) is modified with an ultrafast pulse shaper capable of independent amplitude and phase modulation~\cite{Weiner00}.

The LO spectrum is partitioned in either 4, 6, or 10 bands of equal energy.
By scanning the relative phase between the down-converted comb and the LO, the $x$ and $p$ quadrature noises are measured from the state projected onto the spectral composition of the LO mode.
The quadrature noises are then recorded for each spectral region as well as all possible pairs of regions.
Upon doing so, a covariance matrix is assembled that furnishes a good approximation of the full quantum state.
Cross correlations of the form $\langle \hat{x} \hat{p} +\hat p\hat x\rangle$ are observed to be negligible, which enables the covariance matrix to be expressed in a block diagonal form, i.e., one block for the $x$ quadrature and another for the $p$ quadrature~\cite{ARCFFT14}; cf. Eq.~\eqref{Eq:4}.
From the data contained in the covariance matrix, it is possible to extract special modes, called {\sl supermodes},  that are the eigenmodes of the parametric interaction~\cite{PVTF10} and are uncorrelated with each other.
They turn out to be significantly squeezed~\cite{RMJFT14}, as shown in Table~\ref{tab:Data}.
The existence of squeezed supermodes that span the entire frequency spectrum is at the origin of the entanglement that exists between the frequency bands.

\begin{table}[ht]
	\caption{
		The highest supermode squeezing (sqz.) and antisqueezing levels of the considered 4-, 6-, and 10-mode quantum comb states.
		All of the noise levels are specified in decibels (dB).
	}
	\label{tab:Data}
	\begin{tabular}{rrrrrr}\hline
		\multicolumn{2}{c}{4 modes}	&	\multicolumn{2}{c}{6 modes}	&	\multicolumn{2}{c}{10 modes}		\\
		\phantom{anti}sqz.	&	antisqz.	&	\phantom{anti}sqz.	&	antisqz.		&	\phantom{anti}sqz.	&	antisqz.	\\\hline\hline
		-5.1 dB		&	7.1 dB		&	-2.6	dB		&	3.0 dB		&	-3.7 dB		&	5.8 dB				
	\\\hline
	\end{tabular}	
\end{table}

\section{Data analysis}
\begin{figure}[ht]
	\centering
	\includegraphics[width=85mm]{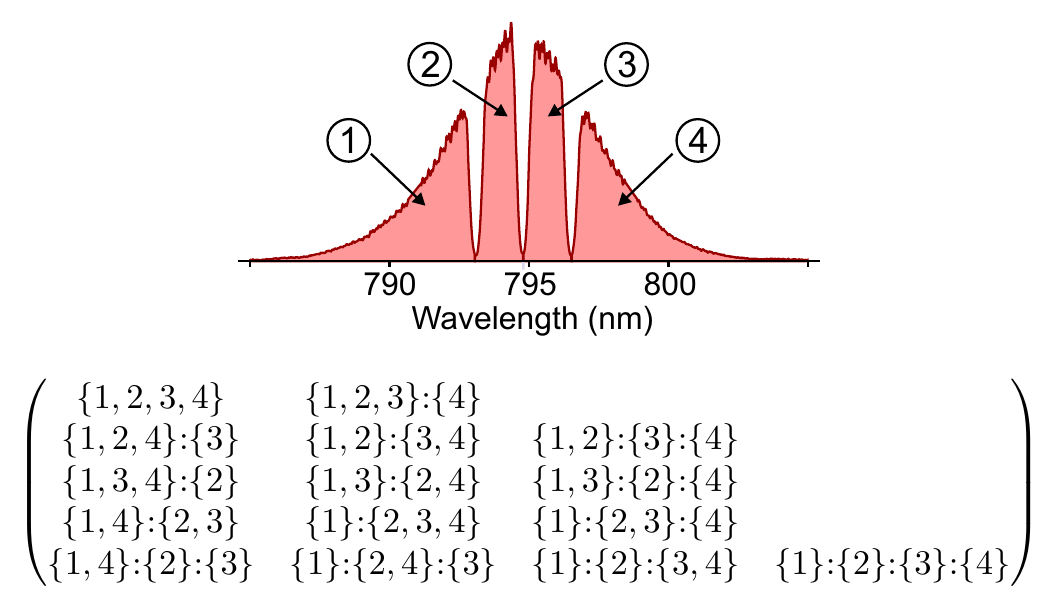}
	\caption{(Color online)
		Structure of 4-mode state. The spectral components (top panel) and partitionings (bottom panel) are shown.
	}\label{Fig:4modestructure}
\end{figure}

The genetic algorithm is implemented for every possible partition of the states $\hat\rho_{N}$ where $N = 4,6, \textrm{and}\,10$.
The mode decompositions $\mathcal{I}_1{:}\cdots{:}\mathcal{I}_K$ are realized with the map	$P:\{1,\ldots,N\}\mapsto\{1,\ldots,K\}$, where $P$ maps $P(j)=k$, if and only if $j\in\mathcal{I}_k$.
Consequently, the $K$ partitions of the original $N$-member set can be arranged in matrix form, which is adapted from the Bell triangle (also referred to as Aitken's array or the Peirce triangle).
The mode labels range from the highest frequency spectral components $1$ to the lowest frequency components $N$ in ascending order.
For example, the mode partitioning, along with the relevant spectral components, for $N=4$ is depicted in Fig.~\ref{Fig:4modestructure}.

Because of to a measurement time ranging between $\sim 10$ and 30 minutes per matrix, slowly varying drifts may render the covariance matrix slightly unphysical.
In order to counter these effects, white noise is added to the experimentally measured covariance matrices so that the minimal symplectic eigenvalue of the noisy matrix becomes positive; cf.~\cite{supplement}.
It is important to emphasize that such a local noise convolution is a separable operation, and therefore is unable to induce entanglement from an originally separable state.
Without this procedure, negative $\Sigma$ values might occur, which are due to a violation of positivity of the covariance matrix instead of being authentic entanglement evidence.
This extra noise is taken into account when obtaining $\Sigma$.

\section{Results}
The results of our methodology for the $4$-mode states are detailed in the matrix $\Sigma_{N=4}$ shown below.
The significances $\Sigma$ are calculated according to Eq.~\eqref{eq:Sigma}, and a particular element in the displayed matrix corresponds to the mode partition shown at the same position of the matrix in Fig.~\ref{Fig:4modestructure},
\begin{align*}
		\Sigma_{N=4}=&\begin{pmatrix}
		    0.01	&	-21.06		&			&	\\
		    -11.21	&	-24.34		&	-24.59		&	\\
		    -13.17	&	-23.52		&	-23.97		&	\\
		    -4.66	&	-20.93		&	-21.63		&	\\
		    -13.16	&	-24.03		&	-24.32		&	-24.61		
		\end{pmatrix}.
\end{align*}
The first entry in the matrix is the trivial partition with only one party, $\mathcal I_1=\{1,\ldots,N\}$, and, therefore, must not exhibit entanglement.
The following 14 partitions, however, are each entangled to a significant degree ($ | \Sigma | \gtrsim 4$, corresponding to a confidence level of $99.99\%$).
The partition displaying the highest entanglement significance is not a bipartition, but rather coincides with the total division of the state into $N$ independent structures.
During the down-conversion process, the initial onset of any quantum correlation among the frequency bands invalidates the full separability of the state.
Thus, this partition is the first to become entangled during down-conversion, and therefore exhibits the most significant entanglement. 
Conversely, the least significantly entangled partition corresponds to detaching the spectral wings (elements $\{1,4\}$) from the spectral center (elements $\{2,3\}$).
This partition indicates an asymmetric distribution of entanglement with respect to the central frequency of the comb. 
In general, symmetric quantum correlations in the comb are stronger since the preponderance of the down-conversion events originate from the pump spectral center.
Asymmetric frequency correlations originate from down-conversion events displaced from the pump central frequency, which therefore occur with lower probability. 
Since the partition $\{1,4\}:\{2,3\}$ demands the highest degree of asymmetric correlations, it possesses a lowered entanglement significance.
Nevertheless, the fact that all of the nontrivial partitions are entangled implies that each resolvable frequency band is entangled with every other band (i.e., the full entanglement of the comb). 
Importantly, this characteristic of the quantum comb would go unnoticed without the use of entanglement criteria capable of probing higher-order state partitions, i.e., $K > 2$.

\begin{figure}[ht]
	\centering
	\includegraphics*[width=8cm]{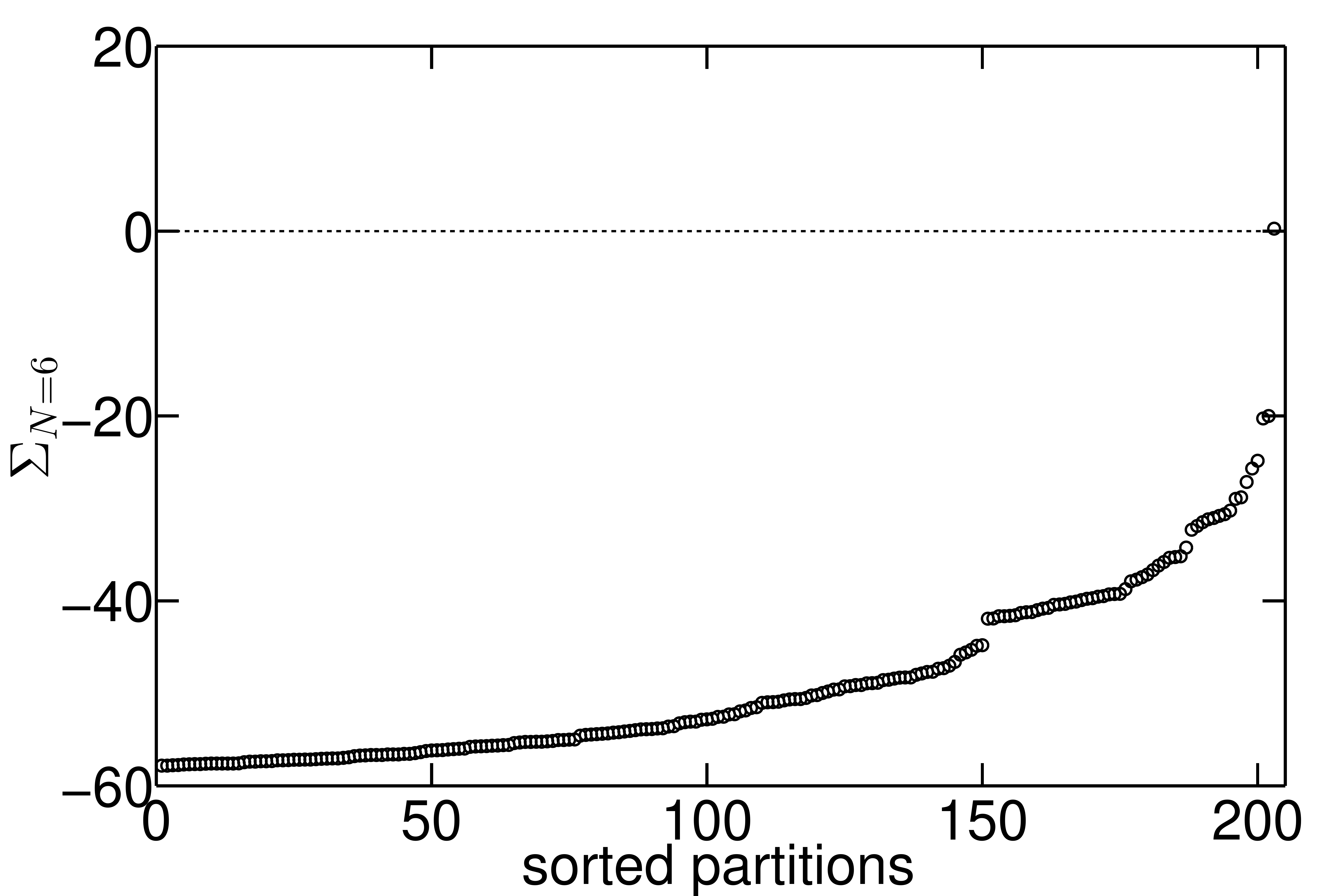}
	\caption{
		Significance of all partitions for the 6-mode states where the partitions are ordered according to the significance of the detected entanglement.
		All of the values are negative except for a single positive value, which represents the trivial partition, $K=1$, and cannot be entangled.
	}\label{Fig:6mode}
\end{figure}

\begin{figure}[ht]
	\centering
	\includegraphics*[width=8cm]{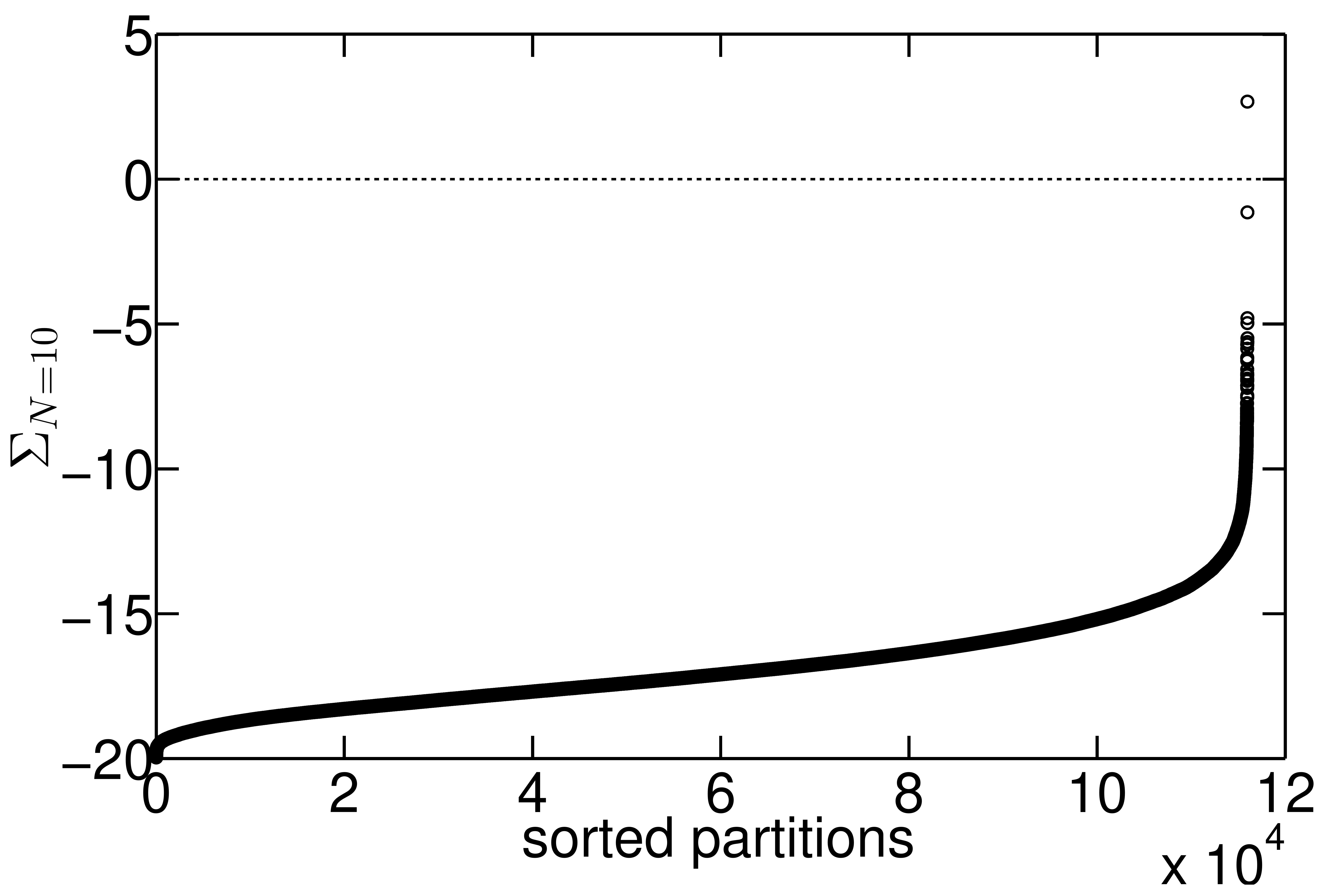}
	\caption{
		The verified entanglement for all 115\,974 nontrivial partitions -- sorted by significance $\Sigma$ -- for the 10-mode frequency-comb Gaussian state.
	}\label{Fig:10mode}
\end{figure}

In the case of 6 modes, 203 unique mode partitions are possible, and the resultant entanglement metric $\Sigma$ is displayed in Fig.~\ref{Fig:6mode}.
The results for the entire set of unique partitions of the 10-mode scenario are likewise depicted in Fig.~\ref{Fig:10mode}; cf. also Ref.~\cite{supplement}.
All of the partitions in both the 6- and 10-mode combs are demonstrated to be entangled except for the trivial partition.

Specific $K$ partitions and their corresponding entanglement metrics $\Sigma$ are shown in Table~\ref{Tab:10mode} for the 10-mode comb state.
Within the $K=2$ subgroup, the most significantly entangled partition results from bisecting the spectrum at its center, whereas the least significantly entangled structure originates from disconnecting the two extreme spectral zones from the remaining spectrum.
This result is consistent with previous observations~\cite{RMJFT14} as well as the results shown above for the 4-mode state.
Additionally, 41\,863 partitions ($\sim 36 \%$) of the 10-mode state reveal an entanglement more significant than that detected for any of the 511 possible state bipartitions.
Hence, a richer understanding of the quantum phenomena implicit in the multimode state is afforded only upon examination of these higher-order state partitions.
As before, the complete dissolution of the frequency comb structure into ten discrete bins is among the most significantly entangled partitions. 

\begin{table}[ht]
	\centering
	\caption{
		The lowest and highest significances of all $K$ partitions, $\mathcal I_1{:}\cdots{:}\mathcal I_K$, for the 10-mode state are given.
	}\label{Tab:10mode}
	\begin{tabular}{lcr}\hline
		$K$ & Partition & $\Sigma$ \\
		\hline\hline
		$1$ &		$\{1,2,3,4,5,6,7,8,9,10\}$ 											& $+2.7$ \\
		$2$ &		$\{1,10\}{:}\{2,3,4,5,6,7,8,9\}$ 										& $-1.1$ \\
		$2$ &		$\{1,2,3,4,5\}{:}\{6,7,8,9,10\}$ 										& $-17.6$ \\
		$3$ &		$\{1,10\}{:}\{2,3,8,9\}{:}\{4,5,6,7\}$ 									& $-5.5$ \\
		$3$ &		$\{1,2,3,4,5\}{:}\{6,9,10\}{:}\{7,8\}$ 									& $-18.9$ \\
		$4$ &		$\{1,10\}{:}\{2,9\}{:}\{3\}{:}\{4,5,6,7,8\}$								& $-8.0$ \\
		$4$ &		$\{1,2,3,4\}{:}\{5\}{:}\{6,9,10\}{:}\{7,8\}$ 								& $-20.0$ \\
		$5$ &		$\{1,10\}{:}\{2\}{:}\{3\}{:}\{4,5,6,7,8\}{:}\{9\}$ 							& $-9.4$ \\
		$5$ &		$\{1,6\}{:}\{2,5\}{:}\{3,4\}{:}\{7,10\}{:}\{8,9\}$							& $-19.8$ \\
		$6$ &		$\{1,10\}{:}\{2\}{:}\{3\}{:}\{4,5,6,7\}{:}\{8\}{:}\{9\}$						& $-11.6$ \\
		$6$ &		$\{1,7\}{:}\{2,5\}{:}\{3\}{:}\{4,10\}{:}\{6\}{:}\{8,9\}$						& $-19.9$ \\
		$7$ &		$\{1,10\}{:}\{2\}{:}\{3\}{:}\{4\}{:}\{5,6,7\}{:}\{8\}{:}\{9\}$					& $-14.3$ \\
		$7$ &		$\{1,5\}{:}\{2,4\}{:}\{3\}{:}\{6,9\}{:}\{7\}{:}\{8\}{:}\{10\}$					& $-19.8$ \\
		$8$ &		$\{1,10\}{:}\{2\}{:}\{3\}{:}\{4,7\}{:}\{5\}{:}\{6\}{:}\{8\}{:}\{9\}$				& $-15.8$ \\
		$8$ &		$\{1,5\}{:}\{2\}{:}\{3\}{:}\{4\}{:}\{6\}{:}\{7,10\}{:}\{8\}{:}\{9\}$				& $-19.7$ \\
		$9$ &		$\{1,10\}{:}\{2\}{:}\{3\}{:}\{4\}{:}\{5\}{:}\{6\}{:}\{7\}{:}\{8\}{:}\{9\}$			& $-16.8$ \\
		$9$ &		$\{1\}{:}\{2,5\}{:}\{3\}{:}\{4\}{:}\{6\}{:}\{7\}{:}\{8\}{:}\{9\}{:}\{10\}$			& $-19.7$ \\
		$10$ &		$\{1\}{:}\{2\}{:}\{3\}{:}\{4\}{:}\{5\}{:}\{6\}{:}\{7\}{:}\{8\}{:}\{9\}{:}\{10\}$ 	& $-19.3$ \\
		\hline
	\end{tabular}
\end{table}

It is worth noting that the number of analyzed frequency bands is currently limited by the optical resolution of the pulse shaper.
A new generation of the setup should allow for observing at least 30 frequency modes as predicted by theory in the present experimental conditions.

\section{Conclusions}
We implemented covariance-based, high-order entanglement criteria on the multimode squeezed states contained within an ultrafast frequency comb.
A genetic algorithm was exploited to maximize the statistical significance of the determined entanglement.
Upon doing so, the criterion identifies entanglement in all of the 14, 202, and 115\,974 nontrivial partitions of the 4-, 6-, and 10-mode scenarios, respectively.
Consequently, the quantum comb exhibits full multipartite entanglement, i.e., entanglement for all partitionings.
Importantly, the currently employed criterion was able to identify entanglement not recognizable with traditional separability metrics.
The present approach allows for the identification of partially and fully entangled states for applications in quantum communication or cluster state computation.

\section{Acknowledgement}
This work has been supported by Deutsche Forschungsgemeinschaft through SFB 652, the European Research Council starting grant Frecquam, and the French National Research Agency project Comb.
C.F. is a member of the Institut Universitaire de France. J.R. acknowledges support from the European Commission through Marie Curie Actions.


\begin{thebibliography}{99}
	\bibitem{EPR35} A. Einstein, N. Rosen, and B. Podolsky, Phys. Rev. \textbf{47}, 777 (1935).
	\bibitem{SB35} E. Schr\"odinger, Proc. Cambridge Philos. Soc. \textbf{31}, 555 (1935).
	\bibitem{SD36} E. Schr\"odinger, Proc. Cambridge Philos. Soc. \textbf{32}, 446 (1936).
	\bibitem{NielsenChuang} M. A. Nielsen and I. L. Chuang, \textit{Quantum Computation and Quantum Information} (Cambridge University Press, Cambridge, England, 2000).
	\bibitem{HHHH09} R. Horodecki, P. Horodecki, M. Horodecki, and K. Horodecki, Rev. Mod. Phys. \textbf{81}, 865 (2009).
	\bibitem{GT09} O. G\"uhne and G. T\'oth, Phys. Rep. \textbf{474}, 1 (2009).
	\bibitem{HV13} M. Huber and J. I. de Vicente, Phys. Rev. Lett. \textbf{110}, 030501 (2013). 
	\bibitem{LM13} F. Levi and F. Mintert, Phys. Rev. Lett. \textbf{110}, 150402 (2013). 
	\bibitem{GK14} G. Giedke and B. Kraus, Phys. Rev. A \textbf{89}, 012335 (2014). 
	\bibitem{Peres96} A. Peres, Phys. Rev. Lett. \textbf{77}, 1413 (1996). 
	\bibitem{Simon00} R. Simon, Phys. Rev. Lett. \textbf{84}, 2726 (2000). 
	\bibitem{DGCZ00} L.-M. Duan, G. Giedke, J. I. Cirac, and P. Zoller, Phys. Rev. Lett. \textbf{84}, 2722 (2000). 
	\bibitem{WW01} R. F. Werner and M. M. Wolf, Phys. Rev. Lett. \textbf{86}, 3658 (2001). 
	\bibitem{DSHPES11} J. DiGuglielmo, A. Samblowski, B. Hage, C. Pineda, J. Eisert, and R. Schnabel, Phys. Rev. Lett. \textbf{107}, 240503 (2011). 
	\bibitem{LF03} P. van Loock and A. Furusawa, Phys. Rev. A \textbf{67}, 052315 (2003). 
	\bibitem{AB05} G. S. Agarwal and A. Biswas, New J. Phys. \textbf{7}, 211 (2005). 
	\bibitem{SV05} E. Shchukin and W. Vogel, Phys. Rev. Lett. \textbf{95}, 230502 (2005). 
	\bibitem{Serafini06} A. Serafini, Phys. Rev. Lett. \textbf{96}, 110402 (2006). 
	\bibitem{HE06} P. Hyllus and J. Eisert, New J. Phys. \textbf{8}, 51 (2006). 
	\bibitem{SV06} E. Shchukin and W. Vogel, Phys. Rev. A \textbf{74}, 030302(R) (2006). 
	\bibitem{HZ06} M. Hillery and M. S. Zubairy, Phys. Rev. Lett. \textbf{96}, 050503 (2006). 
	\bibitem{GHGE07} O. G\"uhne, P. Hyllus, O. Gittsovich, and J. Eisert, Phys. Rev. Lett. \textbf{99}, 130504 (2007). 
	\bibitem{MPHH09} A. Miranowicz, M. Piani, P. Horodecki, and R. Horodecki, Phys. Rev. A \textbf{80}, 052303 (2009). 
	\bibitem{SSV13} F. Shahandeh, J. Sperling, and W. Vogel, Phys. Rev. A \textbf{88}, 062323 (2013). 
	\bibitem{STJZXP07} X. Su, A. Tan, X. Jia, J. Zhang, C. Xie, and K. Peng, Phys. Rev. Lett. \textbf{98}, 070502 (2007). 
	\bibitem{YULF08} M. Yukawa, R. Ukai, P. van Loock, and A. Furusawa, Phys. Rev. A \textbf{78}, 012301 (2008). 
	\bibitem{PMSBF11} M. Pysher, Y. Miwa, R. Shahrokhshahi, R. Bloomer, and O. Pfister, Phys. Rev. Lett. \textbf{107}, 030505 (2011). 
	\bibitem{AMJHTLB12} S. Armstrong, J.-F. Morizur, J. Janousek, B. Hage, N. Treps, P. K. Lam, and H.-A. Bachor, Nat. Commun. \textbf{3}, 1026 (2012). 
	\bibitem{CBCVMN09} A. S. Coelho, F. A. S. Barbosa, K. N. Cassemiro, A. S. Villar, M. Martinelli, and P. Nussenzveig, Science \textbf{326}, 823 (2009). 
	\bibitem{VSEHFS13} C. E. Vollmer, D. Schulze, T. Eberle, V. H\"andchen, J. Fiur{\'a}{\v{s}}ek, and R. Schnabel, Phys. Rev. Lett. \textbf{111}, 230505 (2013). 
	\bibitem{HHH96} M. Horodecki, P. Horodecki, and R. Horodecki, Phys. Lett. A \textbf{223}, 1 (1996). 
	\bibitem{HHH01} M. Horodecki, P. Horodecki, and R. Horodecki, Phys. Lett. A \textbf{283}, 1 (2001). 
	\bibitem{SV13} J. Sperling and W. Vogel, Phys. Rev. Lett. \textbf{111}, 110503 (2013). 
	\bibitem{SV09} J. Sperling and W. Vogel, Phys. Rev. A \textbf{79}, 022318 (2009). 
	\bibitem{RMJFT14} J. Roslund, R. Medeiros de Ar\'ajo, S. Jiang, C. Fabre, and N. Treps, Nat. Photon. \textbf{8}, 109 (2014).
	\bibitem{AI07} G. Adesso and F. Illuminati, J. Phys. A \textbf{40}, 7821 (2007). 
	\bibitem{BT10} D. Berend and T. Tassa, Probab. Math. Stat. \textbf{30}, 185 (2010). 
	\bibitem{CommentTrace} 
		Note that we use the notion ``${\rm tr}$'' for the quantum mechanical trace, e.g., ${\rm tr}(\hat L)$, whereas ``${\rm Tr}$'' corresponds to the trace of a matrix, ${\rm Tr}(M)$ for $M\in\mathbb C^{d\times d}$.
	\bibitem{supplement} See Supplemental Material at \url{http://link.aps.org/supplemental/10.1103/PhysRevLett.114.050501} for analytical solutions, numerical implementation, and data processing, which includes 			Refs.~\cite{SV13,GeneticRef,HY00,SCS99}.
	\bibitem{GeneticRef} R. L. Haupt and S. E. Haupt, {\it Practical Genetic Algorithms}, 2nd ed. (John Wiley \& Sons, Hoboken, NJ, 2004).
	\bibitem{VPTF06} G. J. de Valc\'arcel, G. Patera, N. Treps, and C. Fabre, Phys. Rev. A \textbf{74}, 061801(R) (2006). 
	\bibitem{ARCFFT14} R. Medeiros de Ara{\'u}jo, J. Roslund, Y. Cai, G. Ferrini, C. Fabre, and N. Treps, Phys. Rev. A \textbf{89}, 053828 (2014). 
	\bibitem{Weiner00} A. M. Weiner, Rev. Sci. Instrum. \textbf{71}, 1929 (2000). 
	\bibitem{PVTF10} G. Patera, G. De Valc\'arcel, N. Treps, and C. Fabre, Eur. Phys. J. D \textbf{56}, 123 (2010). 
	\bibitem{HY00} L. Huaixin and Z. Youngde, Int. J. Theor. Phys. \textbf{39}, 447 (2000). 
	\bibitem{SCS99} R. Simon, S. Chaturvedi, and V. Srinivasan, J. Math. Phys. (N.Y.) \textbf{40}, 3632 (1999). 
\end{thebibliography}
\end{document}